\documentclass[conference]{IEEEtran}

\usepackage{xcolor}

\usepackage{amsmath}
\usepackage{tikz}
\usepackage{amssymb}
\usepackage{amsthm}
\usepackage{graphicx}
\usepackage{caption}
\usepackage{algorithm}
\usepackage{algpseudocode}
\usepackage{hyperref}
\usepackage{tcolorbox}
\usepackage[affil-it]{authblk}
\def\BibTeX{{\rm B\kern-.05em{\sc i\kern-.025em b}\kern-.08em
    T\kern-.1667em\lower.7ex\hbox{E}\kern-.125emX}}
\usepackage{booktabs}
\usetikzlibrary{positioning, arrows.meta, fit}

\begin{document}
\renewcommand{\(}{\left(}
\renewcommand{\)}{\right)}
\renewcommand{\[}{\left[}
\renewcommand{\]}{\right]}
\renewcommand{\c}{\mathbf{c}}
\renewcommand{\a}{\mathbf{a}}
\renewcommand{\b}{\mathbf{b}}
\newcommand{\m}{\mathbf{m}}
\renewcommand{\j}{\mathbf{j}}
\newcommand{\n}{\boldsymbol{n}}
\renewcommand{\d}{\mathbf{d}}
\newcommand{\s}{\mathbf{s}}
\newcommand{\bmu}{\bm{\mu}}
\renewcommand{\k}{\mathbf{k}}
\newcommand{\f}{\mathbf{f}}
\renewcommand{\S}{\mathbf{S}}
\renewcommand{\L}{\mathbf{L}}
\newcommand{\Sig}{\boldsymbol{\Sigma}}
\newcommand{\sig}{\bm{\sigma}}
\newcommand{\thet}{\bm{\theta}}
\newcommand{\Nabla}{\bm{\nabla}}
\newcommand{\Eta}{\bm{\eta}}
\newcommand{\Ps}{\bm{\Psi}}
\newcommand{\Lam}{\bm{\Lambda}}
\newcommand{\Del}{\bm{\Delta}}
\newcommand{\oDel}{\bm{{\Delta}}}
\newcommand{\blam}{\bm{\lambda}}
\newcommand{\wblam}{\widetilde{\bm{\lambda}}}
\newcommand{\y}{\boldsymbol{y}}
\renewcommand{\H }{\mathbf{H}}
\newcommand{\z}{\mathbf{z}}
\newcommand{\w}{\mathbf{w}}
\newcommand{\p}{\mathbf{p}}
\newcommand{\W}{\boldsymbol{W}}
\newcommand{\E}{\mathbf{E}}
\newcommand{\R}{\mathbf{R}}
\renewcommand{\r}{\mathbf{r}}
\newcommand{\0}{\mathbf{0}}
\newcommand{\D}{\mathbf{D}}
\newcommand{\Z}{\mathbf{Z}}
\newcommand{\1}{\mathbf{1}}
\newcommand{\F}{\mathbf{F}}
\newcommand{\x}{\boldsymbol{x}}
\renewcommand{\a}{\boldsymbol{a}}
\renewcommand{\w}{\boldsymbol{w}}

\newcommand{\I}{\boldsymbol{I}}
\newcommand{\J}{\mathbf{J}}
\newcommand{\C}{\mathbf{C}}
\renewcommand{\P}{\mathbf{P}}
\renewcommand{\t}{\mathbf{t}}
\newcommand{\A}{\boldsymbol{A}}
\newcommand{\T}{\mathbf{T}}
\newcommand{\U}{\mathbf{U}}
\newcommand{\M}{\mathbf{M}}
\newcommand{\h}{\mathbf{h}}
\renewcommand{\u}{\mathbf{u}}
\renewcommand{\v}{\mathbf{v}}
\newcommand{\Q}{\mathbf{Q}}
\newcommand{\K}{\mathbf{K}}
\newcommand{\q}{\boldsymbol{q}}
\newcommand{\V}{\mathbf{V}}
\newcommand{\X}{\mathbf{X}}
\newcommand{\Y}{\mathbf{Y}}
\newcommand{\B}{\mathbf{B}}
\newcommand{\G}{\mathbf{G}}
\newcommand{\g}{\mathbf{g}}
\renewcommand{\b}{\mathbf{b}}
\newcommand{\e}{\mathbf{e}}
\renewcommand{\i}{\mathbf{i}}
\newcommand{\Tr}[1]{{\rm{Tr}}\left\{#1\right\}}
\newcommand{\EE}[1]{{\rm{E}}\left[#1\right]}
\newcommand{\EEN}[1]{{\hat{\rm{E}}}_N\left[#1\right]}
\newcommand{\EEM}[1]{{\hat{\rm{E}}}_M\left[#1\right]}
\newcommand{\EENM}[1]{{\hat{\rm{E}}}_{NM}\left[#1\right]}

\newcommand{\norm}[1]{\left\|#1\right\|}

\newcommand{\COV}[1]{{\rm{Cov}}\left\{#1\right\}}
\newcommand{\Trr}[1]{{\rm{Tr}}^2\left\{#1\right\}}
\renewcommand{\log}[1]{{\rm{log}}#1}
\renewcommand{\arg}[1]{{\rm{arg}}#1}
\renewcommand{\det}[1]{\left|#1\right|}
\newcommand{\diag}[1]{{\rm{diag}}\left\{#1\right\}}
\newcommand{\Diag}[1]{{\rm{Diag}}\left\{#1\right\}}
\newcommand{\sign}[1]{{\rm{sign}}\left\{#1\right\}}
\renewcommand{\Pr}[1]{{\rm{Pr}}\(#1\)}
\renewcommand{\vec}[1]{{\rm{vec}}\(#1\)}
\newcommand{\mat}[1]{{\rm{mat}}\(#1\)}
\newcommand{\imat}[1]{{\rm{mat}}^{-1}\(#1\)}
\newcommand{\val}[1]{{\rm{val}}\(#1\)}
\newcommand{\rank}[1]{{\rm{rank}}\(#1\)}
\newcommand{\Null}[1]{{\mathcal{N}}\left\{#1\right\}}
\newtheorem{lemma}{Lemma}
\newtheorem{definition}{Definition}
\newtheorem{theorem}{Theorem}
\newtheorem{proposition}{Proposition}
\newtheorem{corollary}{Corollary}
\newtheorem{example}{Example}

\title{Self-Supervised Learning for Covariance Estimation}

\author{Tzvi Diskin}
\author{Ami Wiesel}%
\affil{The Hebrew University of Jerusalem}
\maketitle

\begin{abstract}
    We consider the use of deep learning for covariance estimation. We propose to globally learn a neural network that will then be applied locally at inference time. Leveraging recent advancements in self-supervised foundational models, we train the network without any labeling by simply masking different samples and learning to predict their covariance given their surrounding neighbors. The architecture is based on the popular attention mechanism. Its main advantage over classical methods is the automatic exploitation of global characteristics without
    any distributional assumptions or regularization. It can be pre-trained as a foundation model and then be repurposed for various downstream tasks, e.g., adaptive target detection in radar or hyperspectral imagery.
    
\end{abstract}

\section{Introduction}

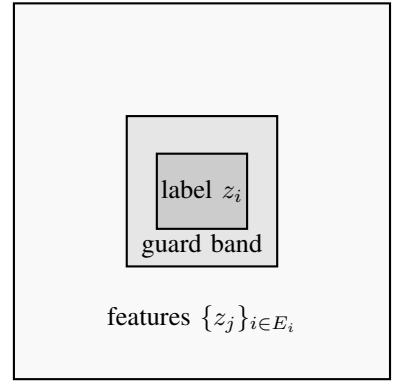
\begin{figure}[t]
    \centering
    \begin{tikzpicture}
        \draw[thick,fill=gray!5] (0,0) rectangle (5,5);
        \draw[thick,fill=gray!20] (1.5,1.5) rectangle (3.5,3.5);
        \draw[thick,fill=gray!40] (1.9,2.0) rectangle (3.1,3.0);
        \node at (2.5,2.5) {label $z_i$};
        \node at (2.5,1.78) {guard band};
        \node at (2.5,0.8) {features $\{z_j\}_{i\in {E_i}}$};
    \end{tikzpicture}
    \caption{Label and feature windows around test sample}
    \label{fig:boxes}
\end{figure}

Covariance estimation is a fundamental problem in statistical signal processing. It is the standard first step before many downstream tasks, e.g. target detection in radar or hyperspectral imagery \cite{manolakis2013detection,nasrabadi2013hyperspectral}. The classical approach is to apply local estimation algorithms around each test sample using its neighboring samples in the outer region after omitting a short band of guard pixels (see Fig. 
\ref{fig:boxes}). The estimation is performed using some variant of maximum likelihood estimation (MLE). There are different types of likelihoods, e.g., Gaussian or Elliptical \cite{zoubir2018robust,sun2016robust,pascal2007covariance,wiesel2015structured}, and assumed structures, e.g., Toeplitz or condition number \cite{babu2016melt,aubry2021new,aubry2017geometric}. The main challenges are accurately choosing between these options and efficiently solving the underlying optimizations. In the last decade, the world began switching to deep learning methods that are ideal for both challenges \cite{shlezinger2023model}. Neural networks are trained on historical data and are optimized to exploit its specific characteristics. In test time, there is no need for any optimization and the data is simply fed into a fixed neural network that efficiently outputs the estimate. Thus, the goal of this paper is to propose a deep learning framework for estimating covariance matrices.

Supervised machine learning requires a large training set with accurate labels. Prior attempts at learning for covariance estimation and target detection problems addressed these challenges using synthetic data with known ground truth covariances \cite{kang2023clutter,barthelme2021doa,addabbo2022nn}. Recently, foundation models have attracted considerable attention. These are trained on large unlabeled datasets via self-supervision. 
The main idea is to mask small subsets of the data and attempt to predict them using the rest of the data. Ideally, this auxiliary task succeeds in learning the underlying statistical properties of the data which can then be used for other downstream tasks. Motivated by these ideas, we propose the Self-Supervised Covariance Estimation (SSCE) framework.  Given a large unlabeled data, we divide it into overlapping pairs of labels and features. The labels are masked samples and the features are the samples in the outer windows surrounding them (see Fig. \ref{fig:boxes}). The framework then uses these pairs to fit a neural network that minimizes a distance function between each label and its predicted covariance. SSCE is based on the popular attention architecture and is designed to directly output the inverse covariance as needed in many downstream tasks. Its main advantage over classical methods is the automatic exploitation of global characteristics without any distributional assumptions or regularization. 
It is suitable for pre-training as a foundation model on large-scale datasets that can then be applied to different and smaller datasets (with or without fine-tuning). 

{\em{Related work:}} SSCE is related to many works in different fields. First, it complements a large body of literature on classical covariance estimation for adaptive target detection \cite{ manolakis2013detection,nasrabadi2013hyperspectral}. In particular, it addresses the same problem as joint estimation of multiple covariance with shared properties \cite{soloveychik2015joint,soloveychik2017joint,besson2007knowledge,bandiera2010knowledge,hua2022unsupervised,JMLR:v16:lee15a}. In the context of learning, SSCE is a deep learning extension of covariance prediction using convex optimization \cite{barratt2023covariance}. Its main novelty with respect to covariance learning using synthetic data \cite{barthelme2021doa,kang2023clutter,addabbo2022nn} is the use of self-supervision. Finally, the main motivation for SSCE is the recent advances in self-supervised foundation models. These are now standard in natural language processing \cite{vaswani2017attention,devlin2018bert,radford2018improving} and are also being developed in other fields closer to SSCE as multispectral imagery \cite{hong2024spectralgpt}. The main novelty of SSCE in this context is that it predicts the second-order statistics of a masked unknown continuous distribution whereas the former predicts discrete tokens out of fixed vocabulary.

\section{Self-Supervised Covariance Estimation}

In this section, we introduce and formulate the Self-Supervised Covariance Estimation (SSCE) framework. Before diving into it, we recall the classical approaches to covariance estimation in a non-stationary environment. 

We consider an environment of $N$ random vectors of dimension $d$, denoted by $z_i$ for $i=1,\cdots,N$.
Their distribution is defined as
\begin{align}
    &z_1,\cdots,z_N|C_1,\cdots,C_N \sim \prod_{i=1}^N p(z_i|C_i)\nonumber\\
    &C_1,\cdots,C_N \sim p(C_1,\cdots,C_N)
\end{align}
where $p(\cdot|C)$ is an unknown zero mean distribution parameterized by a covariance $C\succ 0$, e.g., an Elliptical density. The matrices $C_i$ are
latent variables that characterize the non-stationary environment. Their joint distribution captures their slow variation and the underlying assumption is that
\begin{align}
    i\approx j \quad \Rightarrow \quad C_i\approx C_j .
\end{align}
We emphasize that all the distributions are unknown and the covariances are hidden. We only have access to the dataset $\{z_i\}_{i=1}^N$. 


The goal of this paper is to use the data $\{z_i\}_{i=1}^N$ to efficiently estimate $\{C_i\}_{i=1}^N$. Current local methods for covariance estimation associate each sample $z_i$ with a subset $E_i$ of neighbors (see Fig. \ref{fig:boxes}) and estimate its ${C_i}$ using the samples $\{z_j\}_{j \in E_i}$. The simplest approach is the local sample covariance
\begin{align}
    \hat{C_i}=\frac{1}{|E_i|}\sum_{j\in E_i}z_jz_j^T\qquad i=1,\cdots,N.
\end{align}
More advanced local methods compute $\hat C_i$ using some variant of maximum likelihood estimation:
\begin{align}\label{classic}
\hat C_i =\arg\max_{{C_i}} \sum_{j\in E_i} L(z_j;{C_i})\qquad i=1,\cdots,N.
\end{align}
where $L(\cdot;\cdot)$ is a log-likelihood function. These optimizations need to be computed for each of the $N$ samples and may be computationally expensive. There is also no sharing of global information. Therefore, there are algorithms for joint estimation of all the covariances together with coupling constraints, e.g., a global regularizer $R(\cdot)$ \cite{soloveychik2015joint,soloveychik2017joint,bandiera2010knowledge,hua2022unsupervised,JMLR:v16:lee15a}:
\begin{align}\label{classic2}
\max_{\{{C_i}\}_{i=1}^N} \sum_{i=1}^N\sum_{j\in E_i} L(z_j;{C_i})+ R(\{{C_i}\}_{i=1}^N)
\end{align}

SSCE is the natural next step that replaces these expensive optimizations with a neural network. It is globally trained once on the entire dataset and can then be efficiently applied to any sample under test. Similarly to the classical approach, each predicted ${C_i}$ is a function of $\{z_j\}_{j \in E_i}$. Usually, the challenging part in machine learning is annotating labels, but we do not need any such supervision as we use the sample $z_i$ itself as the label. Self-supervision automatically regularizes the problem and there is also no need for a regularizer. Intuitively, this can be interpreted as a leave-one-out cross validation approach where we leave out $z_i$ and predict it using its neighbors. Altogether, SSCE boils down to solving: 
\begin{tcolorbox}
Self Supervised Covariance Estimation (SSCE)
\begin{align}
\min_{C(\cdot)} \sum_{i=1}^N \ell (z_i, C(\{z_j\}_{j\in E_i})).
\end{align}
\end{tcolorbox}
We choose the loss as the zero mean Gaussian negative log-likelihood:
\begin{align}\label{loss}
    \ell(z;C) = z^TC^{-1}z+\log|C|
\end{align}
However, we emphasize that this is simply a loss function between a label and its prediction. It does not imply any distributional assumption as in (\ref{classic})-(\ref{classic2}).

The architecture of SSCE is based on the self-attention mechanism that is a main building block behind modern foundation models \cite{vaswani2017attention}. Each $z_j$ is embedded into three embeddings: keys $K(\cdot)$, queries $Q(\cdot)$ and values $V(\cdot)$. The embeddings are calculated using a fully connected network with $h$ hidden layers and a ReLU activation.
Then, the samples are mixed using a self-attention layer:
\begin{align}
    &X_l = {\rm Softmax}\left(\frac{Q_l(X_{l-1})  K_l^T(X_{l-1})}{\sqrt{d} }\right)V_l(X_{l-1}) \nonumber\\
    &\qquad\qquad\qquad\qquad\qquad\qquad l = 1,\cdots,L.
\end{align}
 This process is repeated  $L$ times. To ensure positive definiteness, the output inverse covariance is computed as $X_L X_L^T$, where $X_L$ is the output of the last attention layer. Finally, $p$ copies of the network are performed in parallel and their outputs are averaged to get the final inverse covariance. Intuitively, the per-sample deep embedding captures the structure of the inverse covariance matrix, while the self-attention layers aggregate the information and weigh the contribution of each sample to the covariance estimation. Finally, we found empirically that averaging $P$ parallel copies of the model boosts the performance significantly.

 An important property of the SSCE architecture is that it directly models and outputs the inverse covariance, rather than the covariance itself. This is motivated by the fact that the loss function in (\ref{loss}) is only convex in the inverse covariance. It is also useful for downstream tasks such as target detection that also require the inverse. Finally, empirical experiments with and without it show that learning the inverse is much more stable.

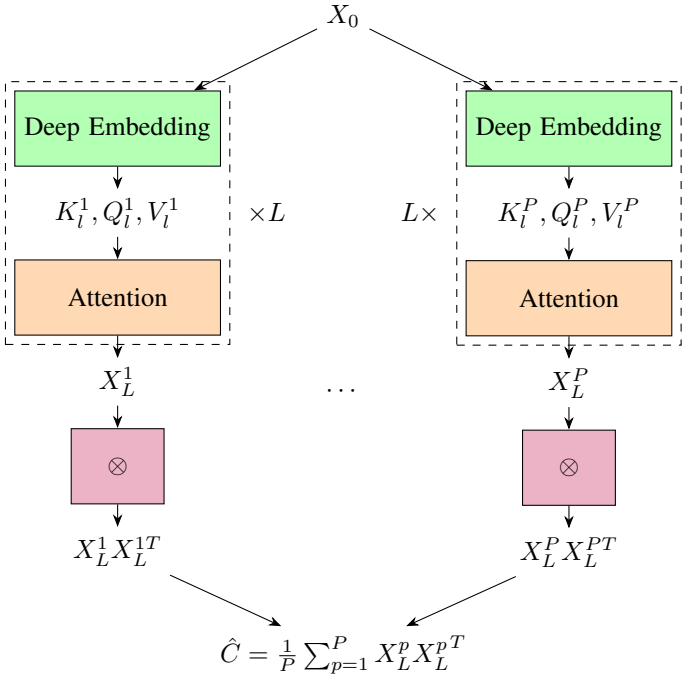
\begin{figure}
  \centering

  \begin{tikzpicture}[
      node distance=0.3cm,
    block/.style={rectangle, draw, text width=2.5cm, text centered, minimum height=1cm},
    arrow/.style={->, >=Stealth},
    embedding/.style={fill=green!30},
    attention/.style={fill=orange!30},
    outer/.style={fill=purple!30,text width=1cm}
  ]
  \node (input)  at (0,0) {${X_0}$};

  \begin{scope}[shift={(-3cm,0)}]
    \node [block, embedding] (hidden11) at (0,-1.5) {Deep Embedding};
    \node [below=of hidden11] (hidden12) {$K^1_l, Q^1_l, V^1_l$};
    \node [block, below=of hidden12, attention] (hidden13) {Attention};

  \node [draw, dashed, fit=(hidden11) (hidden13)] (box) {};
  \node [right=0.1cm of box, align=center] (label) {${ \times L}$};

    \node [below=of hidden13] (hidden14) {${X_{L}^1}$};
    \node [block, below=of hidden14, outer] (hidden15) {$\otimes$};

    \node [ below=of hidden15] (output1) {$X_L^1X_L^{1T}$};

    \draw [arrow] (input) -- (hidden11);
    \draw [arrow] (hidden11) -- (hidden12);
    \draw [arrow] (hidden12) -- (hidden13);
    \draw [arrow] (hidden13) -- (hidden14);
    \draw [arrow] (hidden14) -- (hidden15);
    \draw [arrow] (hidden15) -- (output1);
  \end{scope}

  \node (dots2) at (0,-5) {\textbf{$\cdots$}};

  \begin{scope}[shift={(3cm,0)}]
    \node [block, embedding] (hidden21) at (0,-1.5) {Deep Embedding};
    \node [below=of hidden21] (hidden22) {$K^P_l, Q^P_l, V^P_l$};
    \node [block, below=of hidden22, attention] (hidden23) {Attention};
    \node [below=of hidden23] (hidden24) {${X^P_{L}}$};
   \node [draw, dashed, fit=(hidden21) (hidden23)] (box) {};
  \node [left=0.1cm of box, align=center] (label) {${  L\times}$};
    
    \node [block, below=of hidden24, outer] (hidden25) {$\otimes$};

    \node [ below=of hidden25] (output2) {$X_L^PX_L^{PT}$};

    \draw [arrow] (input) -- (hidden21);
    \draw [arrow] (hidden21) -- (hidden22);
    \draw [arrow] (hidden22) -- (hidden23);
    \draw [arrow] (hidden23) -- (hidden24);
    \draw [arrow] (hidden24) -- (hidden25);
    \draw [arrow] (hidden25) -- (output2);
  \end{scope}

  \node [] (output)  at (0,-8.5) {$\hat{C}=\frac{1}{P}\sum_{p=1}^P X^p_L {X^p_L}^T$};

  \draw [arrow] (output1) -- (output);
  \draw [arrow] (output2) -- (output);

  \end{tikzpicture}
  \caption{Architecture of SSCE}
\end{figure}


\section{Theory}
To shed more light on the advantages of SSCdffdE, we now analyze the asymptotic population case when the empirical loss converges to its true expectation. A seminal result in Bayesian estimation is that the Minimum Mean Squared Estimator (MMSE) that minimizes the $L_2$ loss is the conditional mean:
    \begin{align}\label{mmse}
        \EE{y|x}=\min_{m(\cdot)} \EE{\|y-m(x)\|^2}
    \end{align}
The next theorem generalizes this result to the covariance case\footnote{We note in passing that an alternative loss with the same optimal solution as Theorem 1 is $\EE{yy^T|x}=\min_{C(\cdot)} \EE{\|yy^T-C(x)\|^2}$ which is a special case of (\ref{mmse}). We have experimented numerically with both losses and recommend the log-likelihood loss. It is also easier to extend it to the case of non-zero means. More details will be provided in the journal version.}.
\begin{theorem}
    Let $y$ and $x$ be random variables with a joint distribution $p(x,y)$. Assume $y$ is zero mean and that $\EE{yy^T|x}$ is non-singular for any $x$. Then
    \begin{align}\label{condC}
        \EE{yy^T|x}=\min_{C(\cdot)} \EE{y^TC^{-1}(x)y+\log\left|C(x)\right|}
    \end{align}
\end{theorem}

\begin{proof}
Using the total law of expectation, the objective can be expressed as
    \begin{align}
        &\EE{y^TC^{-1}(x)y+\log\left|C(x)\right|} \nonumber \\
         &    \qquad= \EE{\EE{y^TC^{-1}(x)y+\log\left|C(x)\right||x}} 
    \end{align}
    Solving for each $x$ separately and equating the gradient with respect to $C^{-1}(x)$ to zero yields
    \begin{align}
        \EE{yy^T-C(x)|x}=0
    \end{align}
    Rearranging yields the required result.
\end{proof}

Note the classical MMSE result is distribution-free and holds for any distribution even though the $L_2$ is intuitively associated with Gaussianity. Similarly, although the covariance loss can be interpreted as the Gaussian negative log-likelihood, the theorem is distribution-free and holds for arbitrary non-Gaussian distributions, e.g., Elliptical families that are common in robust covariance estimation. 

\begin{theorem}
Consider an asymptotic SSCE with an expressive architecture, a large training set with $N\rightarrow \infty$ divided into large non-overlapping, independent, and identically distributed (i.i.d.) local environments such that $C_j=C_i$ for all $j\in E_i$ and $|E_i|\rightarrow \infty$. Then, SSCE is consistent and recovers the true unknown covariances
\begin{align}\label{consistent}
C_i^{SSCE}(\{z_j\}_{j \in E_i})\rightarrow C_i.
\end{align}
\end{theorem}
\begin{proof}
Due to $N\rightarrow \infty$ and Theorem 1, we have
\begin{align}\label{Cconv}
C_i^{SSCE}(\{z_j\}_{j \in E_i})\rightarrow\EE{z_iz_i^T|\{z_j\}_{j \in E_i}}.
\end{align}
Due to asymptotic local environments, the posterior converges to
\begin{align}\label{von mises}
p(z_i|\{z_j\}_{j \in E_i})&=\int p(z_i|C_i=C)p(C_i=C|\{z_j\}_{j \in E_i})dC\nonumber\\
&\rightarrow p(z_i|C_i)
\end{align}
where the last step is due to the Bernstein Von Mises Theorem which states that the posterior $p(C_i=C|\{z_j\}_{j \in E_i})$ converges to a Gaussian distribution centered in the consistent maximum likelihood estimator of $C_i$ and its covariance is proportional to $1/|E_i|$ \cite{hartigan1983asymptotic}. Together, combining (\ref{Cconv}), (\ref{von mises}) yields 
\begin{align}\label{Cconv1}
C_i^{SSCE}(\{z_j\}_{j \in E_i})\rightarrow\EE{z_iz_i^T|C_i}=C_i.
\end{align}
\end{proof}

To illustrate SSCE and the two theorems, we consider a classical example of knowledge-aided (KA) covariance estimation where the optimal solution satisfies a simple and intuitive architecture  \cite{besson2007knowledge,bandiera2010knowledge}.  
\begin{example}
 Let $\left\{z_i,\{z_j\}_{j\in E_i}\right\}_{i=1}^N$ be statistically independent pairs of data (corresponding to non-overlapping subsets). Assume the following priors:
\begin{align}
    &z_i, \{z_j\}_{j\in E_i}\;|\; C_i \sim_{\rm{i.i.d.}} {\rm{Normal}}(0,C_i)\nonumber\\
    &C_i \sim_{\rm{i.i.d.}} {\rm{InverseWishart}}(\nu,C)
\end{align}
The SSCE$_{\rm{KA}}$ is implemented with a simple architecture
\begin{equation}
    C_{\rm{KA}}(\{z_j\}_{j\in E_i})=A+\alpha \sum_{j\in E_i}z_jz_j^T
\end{equation}
where $A$ and $\alpha$ are learned global parameters. If $N\rightarrow \infty$ then the optimal solution to SSCE will converge to
\begin{align}\label{asympN}
    &C_{\rm{KA}}(\{z_j\}_{j\in E_i}) \rightarrow \EE{\left.z_iz_i^T\right|\{z_j\}_{j\in E_i}} \nonumber\\
    &\qquad = \EE{\EE{\left.z_iz_i^T\right|C_i,\{z_j\}_{j\in E_i}}|\{z_j\}_{j\in E_i}} \nonumber\\
    &\qquad = \EE{\left.C_i\right|\{z_j\}_{j\in E_i}} \nonumber\\
    &\qquad = \underbrace{\frac{\nu}{\nu+|E_i|-d-1} C}_{A} + \underbrace{\frac{1}{\nu+|E_i|-d-1}}_{\alpha}\sum_{j\in E_i}z_jz_j^T.
\end{align}
Proof: the first equality relies on the law of total expectation and then we used the fact that the conditional distribution of $C_i$ given $\{z_j\}_{j\in E_i}$ is also inverse Wishart with modified parameters.

The results in (\ref{asympN}) are a special case of Theorem 1. They assume $N\rightarrow \infty$ but allow a finite number of neighbors. If we also assume that $|E_i|\rightarrow \infty$ then (\ref{von mises}) holds and SSCE$_{\rm{KA}}$ is consistent:
\begin{align}
    C_{\rm{KA}}(\{z_j\}_{j\in E_i}) \rightarrow \frac{1}{|E_i|}\sum_{j\in E_i}z_jz_j^T \rightarrow C_i.
\end{align}
\end{example}

\section{Numerical examples}
In this section, we evaluate SSCE numerically. We demonstrate the ability of the SSCE to learn an estimator of the covariance without labels and we evaluate its performance using different metrics, including performance on downstream detection and estimation tasks.

We use the architecture described in Section II with the same parameters for all experiments. The fully connected embedding networks include three hidden layers of size 50.The hidden layers of the fully connected layers are of width $50$. The depth of the network is $L=2$ and the number of parallel networks is $P=10$. We train the network for $100000$ iterations with a batch size of $1$. We consider a radar application with complex-valued variables and adopt the natural changes, e.g., switching transposes with conjugate transposes. Similarly to \cite{kang2023clutter}, we use the ComplexPytorch library introduced by \cite{matthes2021learning}. Before the softmax operation we take the absolute value.

We compare SSCE to the following baselines:
\begin{itemize}
    \item RSCM: Regularized sample covariance:
    \begin{equation}
        \hat C_{RSCM} = (1-\alpha) \hat C_{SCM} + \alpha I_d
    \end{equation}
    where $\hat C_{SCM}$ is a local sample covariance and the parameter $\alpha$ is tuned to maximize the performance on the test.
    \item KA: Similarly to Example 1, we estimate the global sample covariance using all the training data together ($\hat C_{GSCM}$) and use it to regularize the local sample covariances:
    \begin{equation}
        \hat C_{KA} = (1-\alpha) \hat C_{SCM} + \alpha \hat C_{GSCM}
    \end{equation}
    \item ATOM - Alternating projection based Toeplitz covariance matrix estimation \cite{aubry2022atom}. 
    \item Oracle: the ground truth covariance matrix (when available).
\end{itemize}

We measure performance using the following metrics:
\begin{itemize}
    \item MSE: average Frobenius norm of the deviation from the true covariance (when available)
    \item NLL: the average negative log-likelihood in (\ref{loss}) evaluated on the test set.
    \item ERR: performance on an estimation downstream task in which we synthetically plant a known target on the data and try to predict its amplitude using the estimated covariance and Weighted Least Squares (WLS). We report the average squared error.
    \item ROC: performance on a detection downstream task in which we synthetically plant a target on the data and try to detect it using the estimated covariance and an Adaptive Matched Filter (AMF) \cite{conte1995asymptotically}. We plot the Receiver Operating Characteristic (ROC) curve and the partial area Under the Curve (AUC) with a maximal FPR of 0.1.
\end{itemize}

\subsection{Synthetic data}
We begin with a synthetic example with non-overlapping, independent and identically distributed local environments. In each local environment, $x_i$ and $\{x_j\}_{j\in E_i}$ are zero mean Complex Normal vectors  with a covariance
\begin{align}
    C_i = \sum_{k=1}^K\sigma^2_{ki}v_kv_k^H + \sigma_w^2I_d   
\end{align}
where $[v_k]_t=e^{j(\omega_k t+\phi_k)}$, $\omega_k$ are 5 fixed frequencies and $\phi_k$ are phases. Different local environments are characterized by different values of $\sigma^2_{ki}$. Specifically, we define 
\begin{equation}
    \sigma_{ki}^2 = A_i [s_i]_k
\end{equation}
where $s$ is generated using Dirichlet distribution with constant parameters $1/10$ and $A_i$ is distributed uniformly $U(0,2)$. This choice yields a sparsity of about 2 active frequencies in each sample. The dimension of the samples here is $d=6$ and the window size is $|E_i|=20$.

\begin{figure}[h]
\centering

\center{\includegraphics[width=0.5\textwidth]{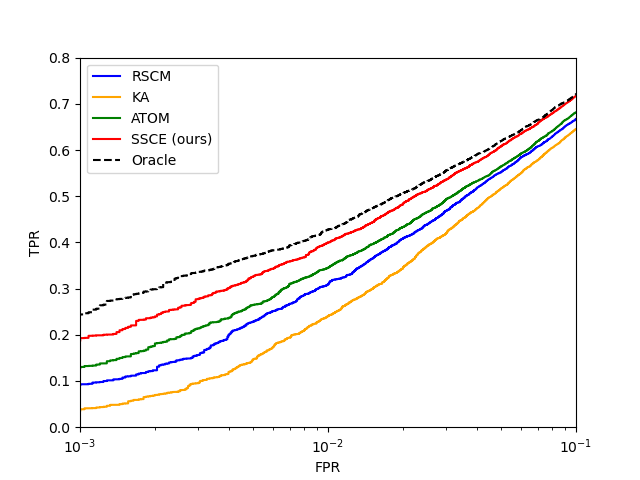}}
\caption{ROC curve of AMF detector in the synthetic data experiment SSCE beats its competitors and is close to the oracle AMF which uses the true covariance matrices}

\label{fig:synthetic_data_roc}%
\end{figure}

Table \ref{synthetic data results} summarizes the performance of the different estimators using the various metrics. SSCE is better than the baselines (except the oracle) in all metrics. Figure \ref{fig:synthetic_data_roc} illustrates the performance gains on the ROC curve. In addition, SSCE is much faster than ATOM, with 

\subsection{IPIX data}
\begin{figure}[h]
\centering

\center{\includegraphics[width=0.5\textwidth]{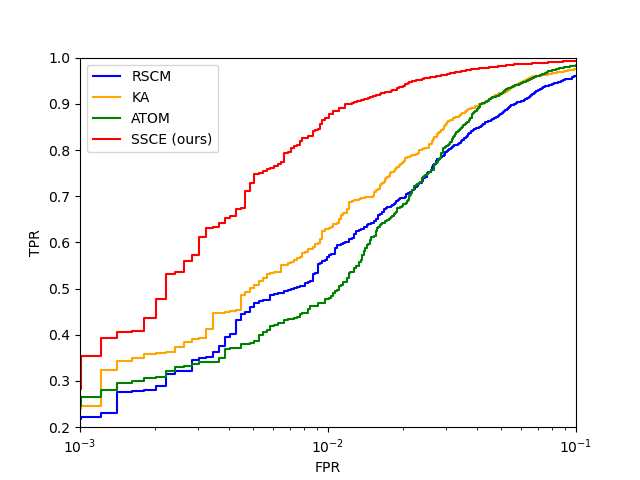}}
\caption{ROC curve of ANMF detector in the IPIX data experiment. SSCE gives better results than its competitors.}

\label{fig:ipix_data_roc}%
\end{figure}

Next, we evaluate SSCE on target detection in IPIX clutter \cite{radar2001mcmaster}. Following \cite{kang2023clutter}, we take samples of length $d=8$ along the time axis and use the $2d$ neighbor (after a single guard cell in each side) range cells as secondary data. We use 14 files for training and 2 different files for testing. In the test set, we synthetically plant targets of the form $[s_k]_t=e^{j(\omega t+\phi_k)}$ with a fixed $\omega$. 
Figure  \ref{fig:ipix_data_roc} compares the ROC curve of the ANMF using the estimated covariance by SSCE versus other baseline covariance estimators. It is easy to see that SSCE is better than all its competitors.

\begin{table}[t]
\caption{Performance metrics in synthetic data}

\label{synthetic data results}
\vskip 0.15in
\begin{center}
\begin{sc}
\begin{tabular}{lcccccr}
\toprule
Detector & MSE & NLL & ERR & AUC \\
\midrule
RSCM & 0.04 & -0.69 & 0.033 & 0.74 \\
KA & 0.04 & -0.19 & 0.031 & 0.72 \\
ATOM & 0.05 & -0.71 & 0.016 & 0.75 \\
SSCE & \textbf{0.02} & \textbf{-1.25} & \textbf{0.014} & \textbf{0.77} \\
\midrule
ORACLE & 0 &  -1.41 & 0.014 & 0.78 \\
\bottomrule

\end{tabular}
\end{sc}
\end{center}
\vskip -0.1in
\label{inference time}

\end{table}

\section{ACKNOWLEDGMENT} 
The authors would like to thank Stefan Feintuch, Joseph Tabrikian for help with IPIX data and Prabhu Babu for providing the code for ATOM. Part of this research was supported by ISF grant 2672/21.

\bibliographystyle{plain}
\bibliography{main.bib}

\end{document}